# Electrical conductivity of cellular Si/SiC ceramic composites prepared from plant precursors

Debopriyo Mallick,[a] Omprakash Chakrabarti,[a)*] Dipten Bhattacharya,[b] Manabendra Mukherjee,[c] Himadri S. Maiti,[b] and Rabindranath Majumdar[d]

[a] *Non-oxide Ceramics and Composites Division, Central Glass and Ceramic Research Institute, Kolkata 700 032, India*
[b] *Electroceramics Division, Central Glass and Ceramic Research Institute, Kolkata 700 032, India*
[c] *Surface Physics Division, Saha Institute of Nuclear Physics, Kolkata 700 064, India*
[d] *Department of Chemical Technology, University of Calcutta, Kolkata 700 009, India*

Electrical conductivity ($\sigma_{dc}$) of the cellular Si/SiC ceramic composites has been measured over a temperature range 25-1073 K while the thermoelectric power (S) has been measured over 25-300 K. Remarkably, these cellular compounds developed through biomimetic route – where the ceramic system grows within a plant bio-template retaining the imprint of structural intricacies of the native templates – are found to exhibit excellent mechanical, thermal, and electrical properties quite comparable to or even better than those of the systems prepared through conventional ceramic route. The electrical conductivity, measured parallel ($\sigma_{\parallel}$) and perpendicular ($\sigma_{\perp}$) to the growth axes of the native plants, depicts nearly temperature-independent anisotropy ($\sigma_{\perp}/\sigma_{\parallel}$) of the order ~2 while the thermoelectric power is nearly isotropic. The charge conduction across the entire temperature regime is found to follow closely the variable range hopping (VRH) mechanism. The conductivity anisotropy appears to be driven primarily by the unique microcellular morphology of the bio-templates which can be exploited in many electrical applications.

*Corresponding author; electronic mail: omprakash@cgcri.res.in

## I. INTRODUCTION

The studies of electrical properties of silicon carbide (SiC) ceramics and composites have tremendous scientific significance and have been proved to be of practical interests as well. SiC materials are long known to be used as high-temperature electric heaters. They are also employed to advantages in resistance thermometers and in thermoelectric power generation targeting various applications such as space technology, local communication, small batteries, refrigeration etc. Such applications involve materials that are developed or in the process of being developed following conventional means of material synthesis using synthetic raw SiC powders. Recently efforts are afoot to synthesize SiC based single-phase or composite materials following an innovative biomimetic route using naturally grown plants or pre-processed bio-structures as precursors. Requirement of cheap plants as precursors of renewable nature and of local source, possibility of easy shape making during green fabrication, possibility of mimicking unique microcellular features of native plant morphology into SiC ceramic microstructures, low temperature ceramization etc. make such processing increasingly attractive for commercial exploitation. Intensive investigations are presently being carried out to examine various aspects of material processing and characterization of biomorphic or cellular SiC materials[1-4]. Although, some preliminary investigations have been undertaken to study the electrical properties of biomorphic SiC ceramics at low[5] as well as at high temperatures[6], no report exists as yet on detailed electrical and thermal conduction mechanisms over a wider temperature range.

In this backdrop, the aim of the present investigation is to study the electrical transport mechanism of cellular Si/SiC ceramic composites, synthesized from bamboo, over a wide temperature range 25-1073 K, for conduction in directions parallel and perpendicular to the growth axis of the precursor bio-template. We have also studied the thermoelectric power over 25-300 K. The local chemical environment has also been studied using x-ray photoelectron spectroscopy (XPS). Our aim is to explore the possibility of application of these cellular Si/SiC composites for designing heaters or power generation circuits with complicated multi-channel network structure having preferential current flow.

Bamboo has been selected as the native plant precursor (or bio-template) for the growth of the ceramic Si/SiC composites, as it has many interesting structural features[7]. Additionally, possibility of value addition to bamboo plants through their successful transformation to SiC based ceramic materials of attractive electrical properties for commercial exploitation, creates an important motive for undertaking the present investigation.

## II. EXPERIMENTAL

### *2.1. Material synthesis*

Cellular Si/SiC ceramics were prepared from bamboos of Indian origin following the procedure described elsewhere.[4] The selected specimen (species: Dendrocalamus Stricuts) is the most common bamboo in India; the specimen used in the present study is of the Western Ghat origin. Precursor samples were collected from lower part of the culms, which were erect and of nearly solid cylindrical forms (major constituents (wt %): α-cellulose= 34.5%, Hemicellulose= 20.5%, lignin=26.0%, Moisture =8.0%, Ash=7.4% ($SiO_2$=6.66%)) Cylindrical shapes (of diameter 30.5 mm and, height 45 mm.) were cut from native bamboo plants, dried at 70°C for 48 h, pyrolyzed by heating at around 800°C in a self–generated atmosphere in an electrically heated furnace followed by heat treatment at around 1600°C under vacuum for 4 h in a carbon tube furnace and finally infiltrated and reacted with molten Si (99.4% w/w) under vacuum at 1600°C for 0.5 h in the same carbon tube furnace to yield cellular Si/SiC ceramics.[4] The bio-carbonaceous preforms from native bamboo were characterized in terms of pyrolysis weight loss, shrinkage, and microstructure; the Si-infiltrated biocarbon preforms were tested for density, porosity, presence of crystalline phases, phase transitions, and also microstructure.

### *2.2. Measurement of electrical properties*

Measurement of low temperature electrical resistivity was carried out from 25 to 300 K by standard four-probe dc technique in a closed cycle cryocooler. The bar shaped specimens of width 2.1-2.6 mm and thickness 1.6-1.8 mm were used. Measurement of thermoelectric power (TEP) was carried out over 25-300 K. For TEP-measurement, two Pt-Pt/Rh thermocouples were attached to both ends of the bar-shaped samples and one end was heated by a Mn-wire heater to generate temperature gradient of the order 3-5 K;

the TEP is obtained from the slope of the thermoelectric force to the temperature difference.

High temperature electrical resistivity was measured over 300-1073 K in a tubular furnace under flowing argon. In this case also four-probe technique was employed for the measurement. High quality Pt wire and paste was used for making the voltage probes and current leads. All the contacts were cured overnight at ~1273 K. The samples were cut into small bars with 3.0-3.3 mm width and 3.0-3.2 mm thickness. Measurements of all the electrical properties were made in directions parallel and perpendicular to growth axis of native bamboo precursors.

***2.3. X-ray photoelectron spectroscopy (XPS) study***

The XPS core-level spectra were taken with an Omicron Multiprobe (Omicron nanotechnology GmbH., UK) spectrometer fitted with an EA 125 hemispherical analyzer. A monochromated Al-K$\alpha$ x-ray source operating at 150W was used for the experiments. The analyzer pass energy was kept fixed at 20 eV for all the scans. In order to remove the surface contaminations, the samples were sputtered with 3 keV argon ions for 30 min prior to the measurements. For quantitative analysis, background of the data was removed using Shirley method and FHI sensitivity factors were used for the calculations.

## III. RESULTS AND DISCUSSION

### *3.1. Materials property evaluation and microstructure characterization of cellular Si/SiC*

Bamboo has many distinguishing structural features[7] — on the macro-scale bamboo is cylinder and in the present case, it is a solid cylinder; on the meso-scale it has a non-linear gradient structure comprising of vascular bundles oriented along the growth axis of bamboo plant thin-walled cells, the walls of which are formed by biopolymers such as cellulose, lignin etc; on the micro-scale bamboo bust fibers are hollow tubes composed of several concentric layers and each layer is reinforced with helically wound cellulose micro fibrils. During pyrolysis, decomposition of biopolymers takes place causing loss in weight and dimensional shrinkages and the results are presented in Table I. Skeletal carbon preforms retains structural integrity and micro-cellular features of native bamboo plant as shown in Fig. 1(a,b); the microstructure is highly anisotropic — in transverse direction to the growth axis, multiplicity of hollow channels of varying diameters originating from tracheidal pores is visible; tubular channel morphology is seen to be preserved in longitudinal direction to the growth axis. Morphological anisotropy appears to have caused difference in pyrolytic shrinkage in different directions (Table I).

Spontaneous infiltration of the porous skeletal carbon preform derived from bamboo plant took place when it was brought into contact with silicon melt under vacuum. Carbon preform got fully infiltrated and reacted throughout into dense structure with complete retention of the macroscopic structural integrity. In the case of Si infiltrated specimen the shrinkages were found to be very low and varied marginally in directions parallel and perpendicular to plant growth axis. Si-infiltrated pyrolyzed bamboo exhibited low density with negligibly small porosity. The material property data are presented in Table I. The XRD-scan from a Si infiltrated pyrolyzed bamboo is shown in Fig. 2. β-SiC and Si are the major crystalline phases present. Additional presence of α-SiC and carbon can also be noticed.

Assuming (i) complete infiltration of Si melt into the systemic pores of the carbon preforms and stoichiometric reaction between C and Si, (ii) no dimensional change and (iii) no loss of materials during ceramization, the material properties can be determined by the following empirical equation. For ideally dense duplex Si/SiC ceramic composites, the ceramic density can be given as:

$$\mathbf{d_{Ceram}} = \mathbf{V_{SiC}}\,\mathbf{d_{SiC}} + \mathbf{V_{Si}}\mathbf{d_{Si}} = \mathbf{d_{Si}} + 1.038\,\mathbf{d_{CB}}(\mathbf{d_{SiC}} - \mathbf{d_{Si}}) \qquad (1)$$

where $\mathbf{V_{SiC}}$ and $\mathbf{V_{Si}}$ are the fractional volumes of SiC and Si present in the duplex Si/SiC ceramic composites and $\mathbf{V_{SiC}} = (d_{CB}M_{SiC})/(d_{SiC}M_C) = 1.038\, d_{CB}$ , $\mathbf{d_{CB}}$ and $\mathbf{M_C}$ being the density of carbon preform and molecular weight of carbon (12 g/ mole), $\mathbf{d_{SiC}}$ and $\mathbf{M_{SiC}}$ the density and molecular weight of SiC (40 g/ mole) respectively and $d_{Si}$ the density of Si. For an average carbon density of 0.47 gm/cc (as measured for bamboo carbon), equation (1) gives a ceramic density of 2.759 gm/cc. Based on this value and the ceramic density obtained in the present study, the relative density can be computed to be 94.95 %, predicting a porosity value of 5.05%. The difference between the predicted and experimental values of porosity may be due to loss of some carbon caused by biopolymeric decomposition during Si infiltration run. Ignoring the presence of pores and other phases in the duplex Si/SiC ceramics, the volume % of Si and SiC-phases may be computed from the experimental density value as 67 and 33% respectively.

During microscopic examination under reflected light microscope (RLM), Si melt infiltrated pyrolyzed bamboo sample showed that the microstructural features of the initial native preform were well preserved during transformation into ceramic structures (Fig. 1c,d). The appearance of dense microstructure was common in both the longitudinal and transverse direction to growth axis. During investigation under SEM, the presence of the major phases of Si and SiC could be confirmed by EDX analysis. Occasional presence of a third phase as deep black spots could be noticed which was likely to be pores and/or residual carbon. In the direction transverse to the growth axis, the pyrolyzed bamboo originally showed tracheidal porous channels with diameter in the range of 2 to 50 μm and such pores are seen to be nearly completely filled with solidified Si after processing. The carbon of the pyrolyzed preform converts to SiC and the residual Si fills the pore interiors. In longitudinal direction to the growth axis, tubular channels were seen to be filled with Si with the cell wall converted to SiC.

### *3.2. Electrical properties of cellular Si/SiC*

A systematic investigation has been carried out to examine the dc resistivity of cellular Si/SiC ceramic composites synthesized from bamboo plant, with an aim to understand the mechanism of charge transport. The variation of dc resistivity as a function of temperature across a range 25-1073 K has been plotted in ln(ρ/T) versus ($1/T^{\gamma}$) format (where γ is related to the dimensionality) in Fig. 3 for directions longitudinal and transverse to the growth axis of native plant. All the curves show more or less similar trends — electrical conductivity increases with temperature. *The value of electrical resistivity at any temperature is comparable to what is reported in Si/SiC system prepared through ceramic route*.[8] There are few interesting features in the entire set of resistivity vs. temperature patterns: (i) nearly temperature-independent anisotropy in resistivity $\rho_{\parallel}/\rho_{\perp} \sim 2$; (ii) cross-over in ρ vs. T pattern at distinct points (T*) where no signature of phase transition could be observed by differential thermal analysis (DTA) study; (iii) compliance with variable range hopping (VRH) scenario as against normal small polaron hopping scenario; (iv) T* is ~256 K for longitudinal direction while ~162 K for transverse direction. The positive thermoelectric power (S) signifies that the charge carriers are holes. The hole concentration does not change much depending on the direction of charge conduction with respect to the growth axis of the bio-template. We attempt to rationalize these features in the light of microstructures of the cellular Si/SiC system.

It is interesting to note that the ρ versus T patterns conform to Mott's VRH scenario where ρ is given by[9]

$$\rho = \rho_0 T.\exp\left(\frac{T_0}{T}\right)^{\gamma} \qquad (2)$$

where the exponent $\gamma$ is related to the dimensionality (d) of the system by $\gamma = (d+1)^{-1}$ relation; $\rho_0$ and $T_0$ are constants which, within the framework of VRH model, are given by

$$\frac{1}{\rho_0} = \sigma_0 = e^2 R^2 \upsilon_{ph} N(E_F) \tag{3}$$

$$T_0 = \frac{\lambda \alpha^3}{k_B N(E_F)} \tag{4}$$

$$R = \left( \frac{9}{8\pi\alpha k_B T N(E_F)} \right)^{\frac{1}{4}} \tag{5}$$

where $\lambda$ is a constant, $\alpha$ (= $1/r_p$) is the inverse of radius of the polaron ($r_p$), $N(E_F)$ is the density of states at the Fermi level, $\upsilon_{ph}$ is the phonon frequency, and R is the average hopping distance. The activation ($E_A$) [Ref. 10] and the hopping energies (W) are given by

$$E_A = k_B \left[ \frac{d \ln\left(\frac{\rho}{T}\right)}{d\left(\frac{1}{T}\right)} \right] . T^{1-\gamma} \tag{6}$$

$$W = \frac{3}{4\pi R^3 N(E_F)} \tag{7}$$

Fitting of the experimentally observed $\rho$ versus T patterns with Eq. (2) over a certain temperature range, yields the parameters $E_A(T)$, W, $r_p$, R, and $N(E_F)$ for specific temperature range as well as for both the directions of the current flow with respect to the growth axis of the bio-template. The values of these parameters have been listed in Table-II; $\lambda$ is taken to be 18.1 while $\upsilon_{ph}$ is considered $10^{15}$ [Ref. 11]. The temperature dependence of the activation energy $E_A(T)$ is shown in Fig. 4. The fitting of the $\rho$ versus

T patterns is shown in Fig. 3. In Figs. 3a-d, we show the fitting with simple small polaron conduction [$\rho = \rho_0 T\exp(E/k_B T)$] and VRH conduction models with $\gamma$ = ½, 1/3, and ¼. Close comparison reveals that ρ versus T patterns – for both longitudinal and transverse directions – fit precisely the 3D-VRH model with $\gamma$ = ¼. In the case of other models, fitting can be done only over a narrow temperature range and, therefore, several crossovers could be noticed. This indicates that these models are not valid for explaining the observed ρ versus T patterns across the entire temperature range. In the case of VRH model with $\gamma$ = ¼, on the other hand, only a single crossover at $T^*$ could be noticed. *This is an important observation*.

We now turn our attention to the thermoelectric power (S) versus temperature patterns. Along both the directions S is found to be increasing with temperature over 25-300 K. Since, Si/SiC composite exhibits semiconducting resistivity pattern over 25-300 K, we employ the standard Heike model in order to estimate the charge carrier concentration (c) participating in heat conduction. According to the model, S is given by[12]

$$S = \frac{k_B}{e} \ln\left[\frac{1-c(T)}{c(T)}\right] \tag{8}$$

In Fig. 5, we show the variation in c(T) with S. Expectedly, in the both cases the variation is found to be nearly identical. The fraction of the charge carriers participating in the heat conduction is progressively decreasing with the increase in temperature. As a result, with the increase in temperature the thermoelectric power is increasing. At room temperature, the charge carrier concentration is estimated to be $3.96\times10^{22}$ $cm^{-3}$ and $4.03\times10^{22}$ $cm^{-3}$ for longitudinal and transverse directions, respectively.

The study of microstructure both in directions longitudinal and transverse to the template growth axis reveals that Si/SiC composite forms a bit complicated network structure, retaining the cellular array feature of the template (Fig. 1). The fibrous wire type of structure along the growth axis is connected through multiple channels along the transverse direction. The current carrying cross-section differs between longitudinal and transverse directions as the cross-sectional area of each of the tubes of the vascular bundle along the growth axis is always smaller than the net surface to surface contact area perpendicular to the growth axis. This current carrying cross-sectional anisotropy which results from morphological anisotropy influences the electrical conduction and gives rise to the anisotropy of nearly 2.

We have examined the local chemical environment along and perpendicular to the growth axis by x-ray photoelectron spectroscopy (XPS). In Fig. 6, we show the corresponding spectra. In both the directions, the system is composed of Si, SiC, and C; some oxygen is also present in association with Si and C.[13] The atom fraction (%) of Si and C in the constituent phases is varying from 44-24-0-32 and 0-19-50-32 to 34-22-0-44 and 0-16-52-32 for longitudinal to transverse directions respectively. Therefore the composition basically appears to be nearly similar in both the directions. The resistivity anisotropy results primarily from morphological anisotropy.

It is interesting to note that ρ vs. T follows 3D-VRH model with $\gamma = 1/4$ as against conventional band semiconductor or small polaron conduction scenario. The VRH model is normally found applicable in – structurally or electronically – disordered systems.[9] In Si/SiC composites, prepared through conventional route, small polaron hopping conduction describes the resistivity patterns.[8] Quite apparent then, that because of network type structure composed of strands of different dimensions and interfaces between remnants of template and Si/SiC, structural disorder develops in cellular Si/SiC systems which gives rise to a switch in charge conduction mechanism. The small polaron parameters such as polaron radius and hopping distance (Table-II) are found to be consistent with those estimated for other such disordered electronic systems.[14-17]

It has been observed that there is a crossover in resistivity pattern at a certain temperature T*; T* is ~162 and 256 K for directions transverse and longitudinal to the template growth axis. The transition appears to be gradual as opposed to sharp. Low temperature DTA study across T*, with a detailed structural analysis can reveal the reason behind the crossover. This is beyond the scope of the present paper and will be attempted in a future work.

## IV. SUMMARY

In summary, we report the electrical resistivity and thermoelectric power of the cellular Si/SiC composites developed from bio-templates over a wide temperature range. The values of these parameters are in excellent agreement to what has been observed in such composites developed from conventional ceramic processes. The electrical resistivity is found to be moderately anisotropic (with anisotropy ~2) whereas the thermoelectric power turns out to be isotropic. The electrical conduction mechanism appears to be following variable range hopping (VRH) scenario both along the direction of growth of the bio-template and perpendicular to it. Because of easy maneuverability of the bio-template structures, complicated designs of electrical heater or power generators for electronic circuitry can be developed using such cellular Si/SiC composites developed via biomimetic routes.

## ACKNOWLEDGMENTS

We thank P. Mandal of Saha Institute of Nuclear Physics, Kolkata, India for the low temperature resistivity and thermoelectric power measurements. We also thank Director, CG&CRI for giving permission to publish this paper. This work is supported by Department of Science and Technology (DST), Government of India (fund sanction no. SR/S3/ME/20/2003-SERC-Engg.).

Table 1. Characteristics of pyrolyzed bamboo and cellular Si/SiC-bamboo

| Characteristics of pyrolyzed bamboo | | | | Characteristics of cellular Si/SiC-bamboo | | | |
|---|---|---|---|---|---|---|---|
| Pyrolytic wt. loss (%) | Linear dimensional shrinkage (%) | | Density ($g.cm^{-3}$) | Density ($g.cm^{-3}$) | Porosity (vol. %) | Linear dimensional shrinkage (%) | |
| | Parallel growth axis | Perpendicular to growth axis | | | | Parallel to growth axis | Perpendicular to growth axis |
| 72.30 | 15.83 | 27.70 | 0.47 | 2.62 | 0.56 | 0.95 | 0.33 |

Note:
1. Carbon density has been determined by measuring weight and linear dimension.
2. Ceramic density has been determined by water displacement method & ceramic porosity by boiling water method.

Table 2. Different parameters associated with charge transport in cellular Si/SiC-bamboo, following 3D-VRH model

| Direction for resistance measurement | Cross over temperature (K) | Temp. Segment (K | Parameters associated with charge transport | | | | |
|---|---|---|---|---|---|---|---|
| | | | $T_o \times 10^5$ (K) | Density of states at Fermi level, $N(E_F)$ ($eV^{-1}cm^{-3}$ | Polaron radius ($r_p$) (Å) | Hopping distance between adjacent sites (Å) | Hopping Energy(eV) |
| Parallel to plant growth axis | 162 | 28-162 | 0.24 | $1.77\times10^{23}$ | 3.66 | 5.48 | 0.008 |
| | | 162-1075 | 7.06 | $1.76\times10^{22}$ | 2.56 | 5.59 | 0.078 |
| Perpendicular to plant growth axis | 256 | 28-256 | 0.27 | $7.32\times10^{23}$ | 2.19 | 3.06 | 0.001 |
| | | 256-983 | 3.46 | $5.29\times10^{21}$ | 4.86 | 8.85 | 0.065 |

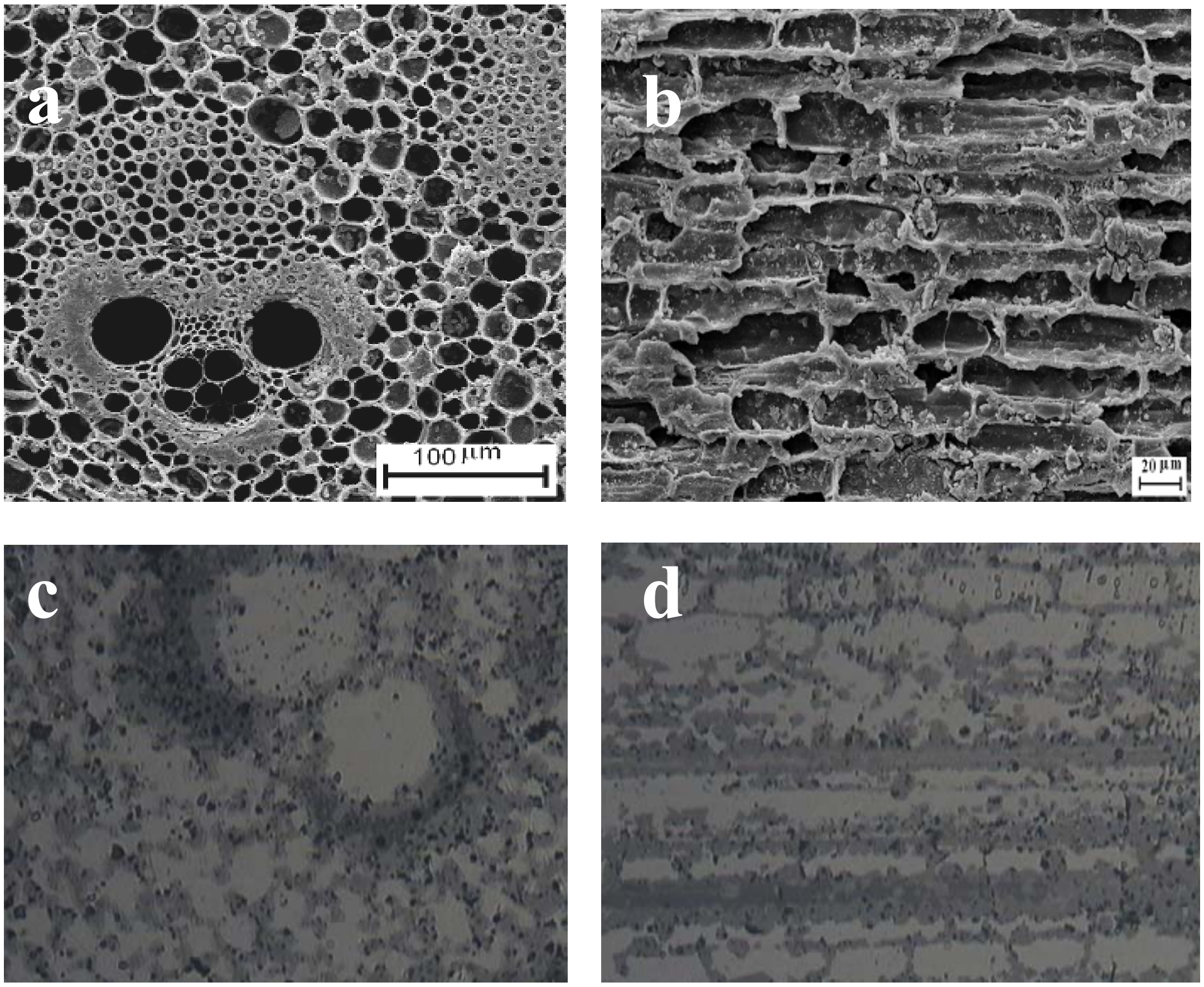

Fig. 1. Preservation of cellular microstructural features of native bamboo plants in the pyrolyzed biocarbon preform (a & b)  and cellular SiSiC ceramic composite (c & d); SEM/SE (a & b) and optical (× 200; c & d) micrographs cross showing multiplicity of hollow channels in transverse sections (a & c) and retention of tubular elongated cell structures in longitudinal sections (b & d) relative to the direction of growth axis of native plants.

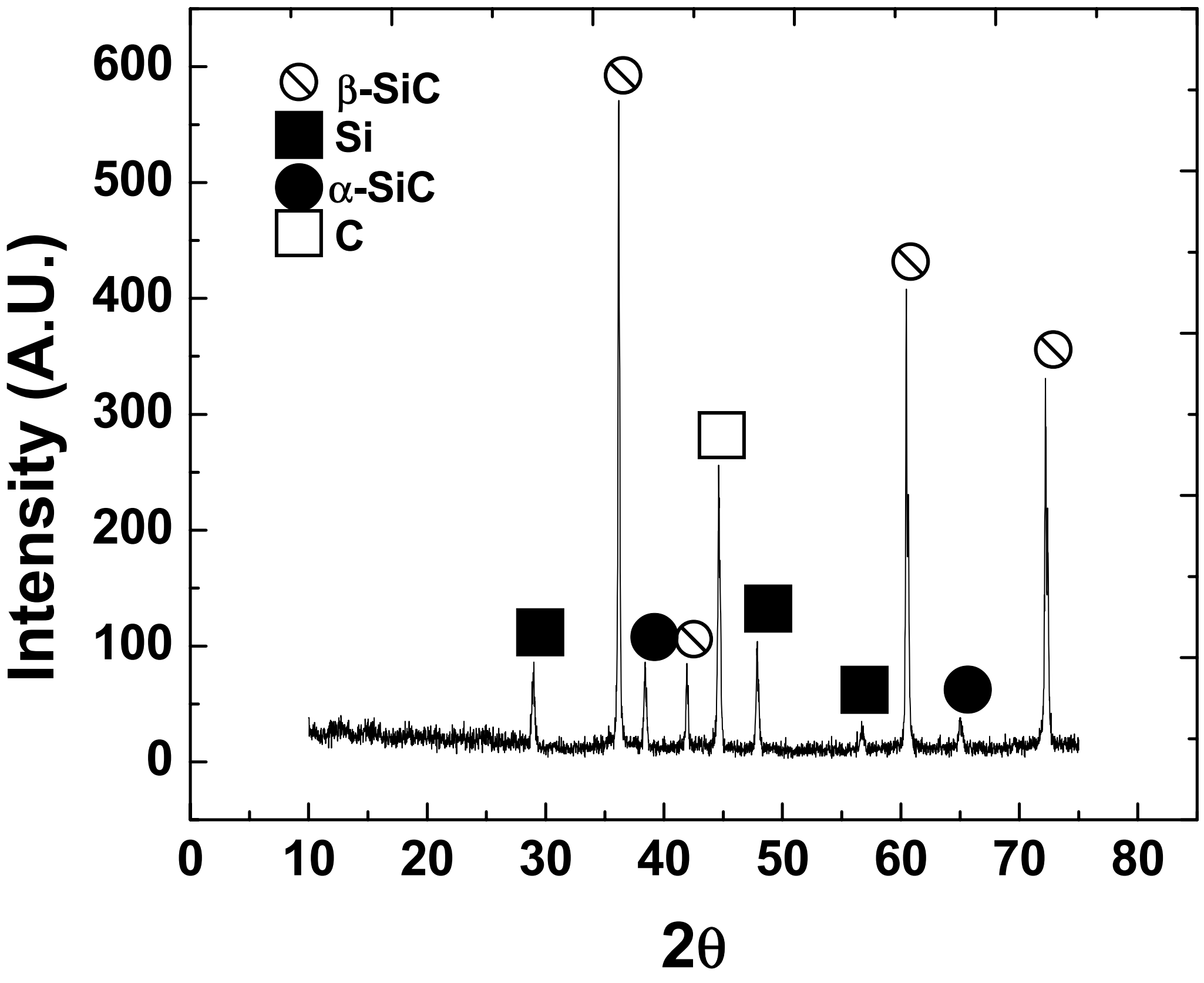

Fig.2. XRD scan of cellular Si/SiC-bamboo showing the presence of Si and β-SiC as the major crystalline phases.

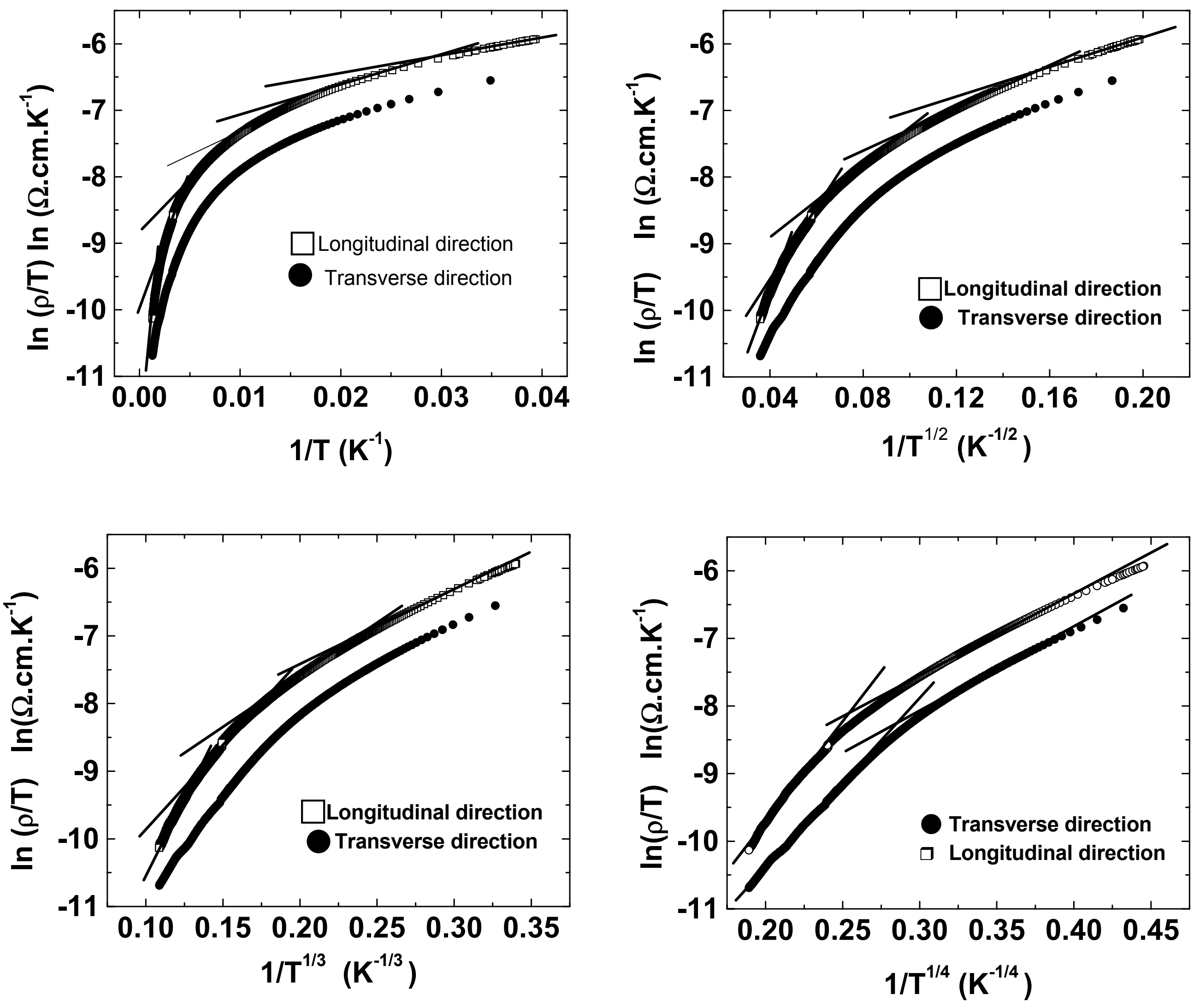

Fig.3. Fitting of experimentally observed ρ versus T patterns following different models: (a) simple small polaron transport; (b) 1D VRH ; (c) 2D VRH; and (d) 3D VRH. It is clear that as we proceed from small polaron to 3D VRH model, the temperature range, over which the fitting is valid, expands which points out the relevance of 3D VRH model in explaining the electrical conduction in cellular Si/SiC composites both in directions longitudinal and transverse to the growth axis of the plant template.

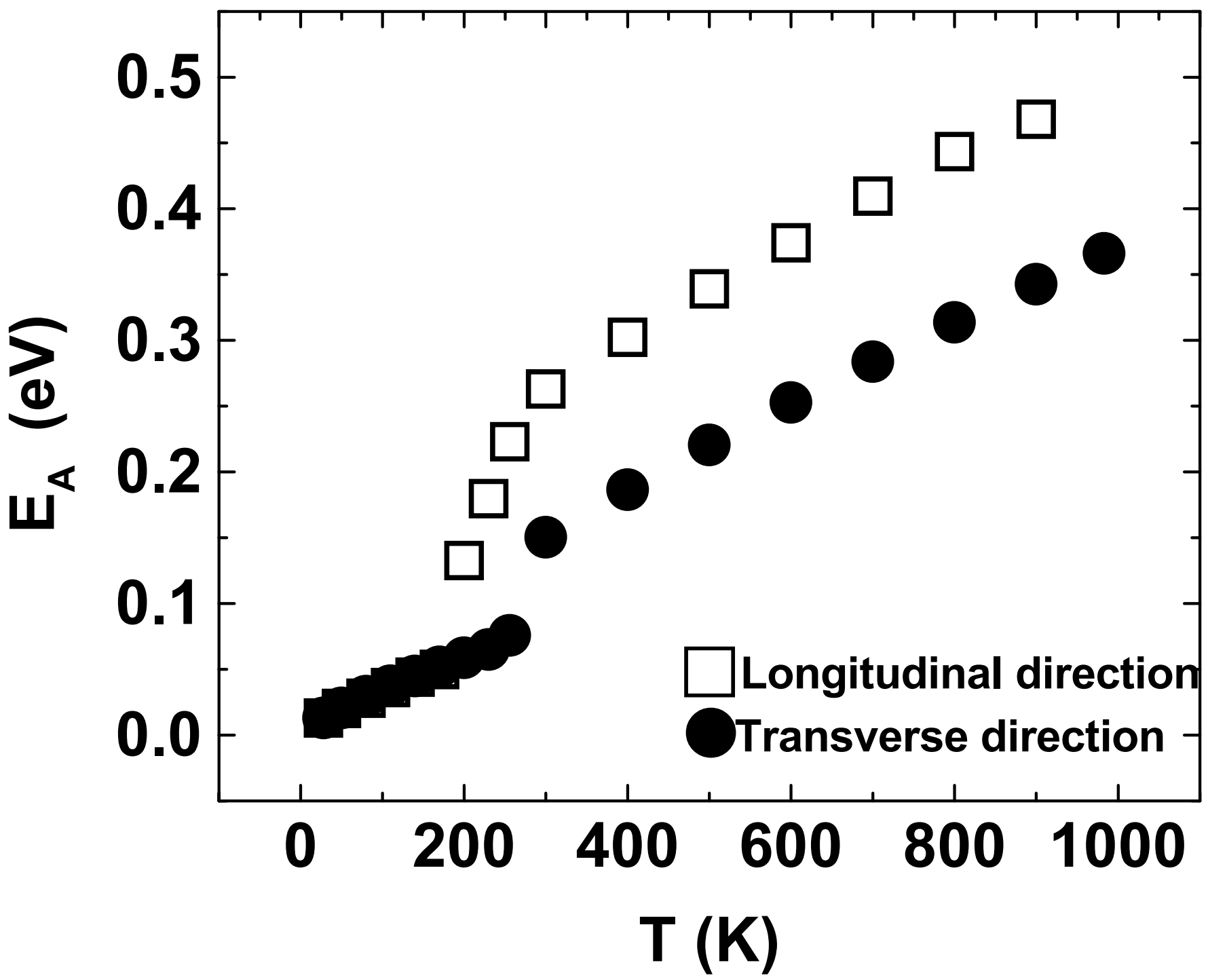

Fig.4. Variation of activation energy ($E_A$) with temperature as estimated from Eq. (6).

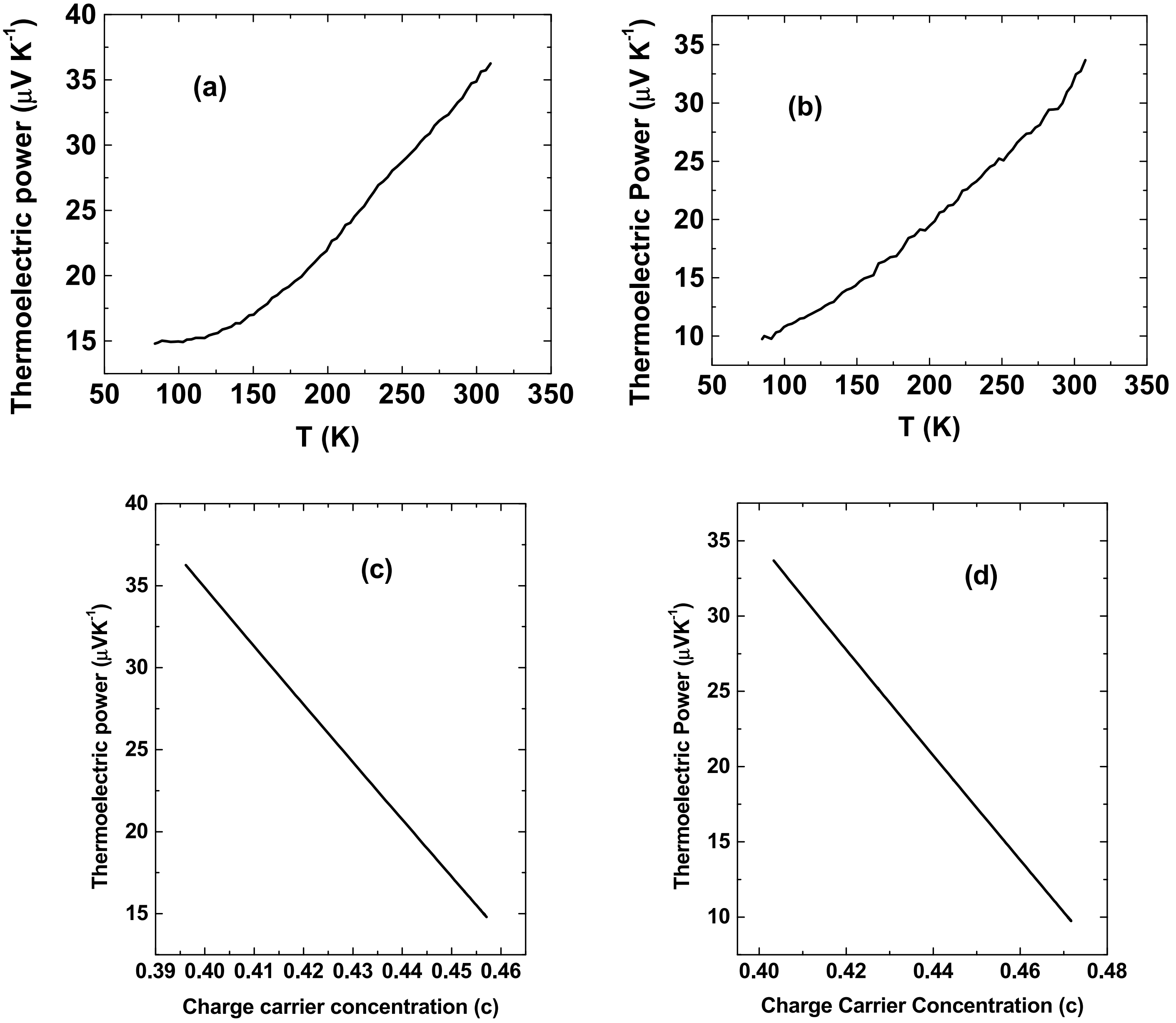

Fig.5. Temperature dependence of thermoelectric power of cellular Si/SiC ceramic composites synthesized from bamboos in direction parallel (a) and perpendicular (b) to the plant growth axis; variation in charge carrier concentration with thermoelectric power in directions parallel (c) and perpendicular (d) to plant growth axis.

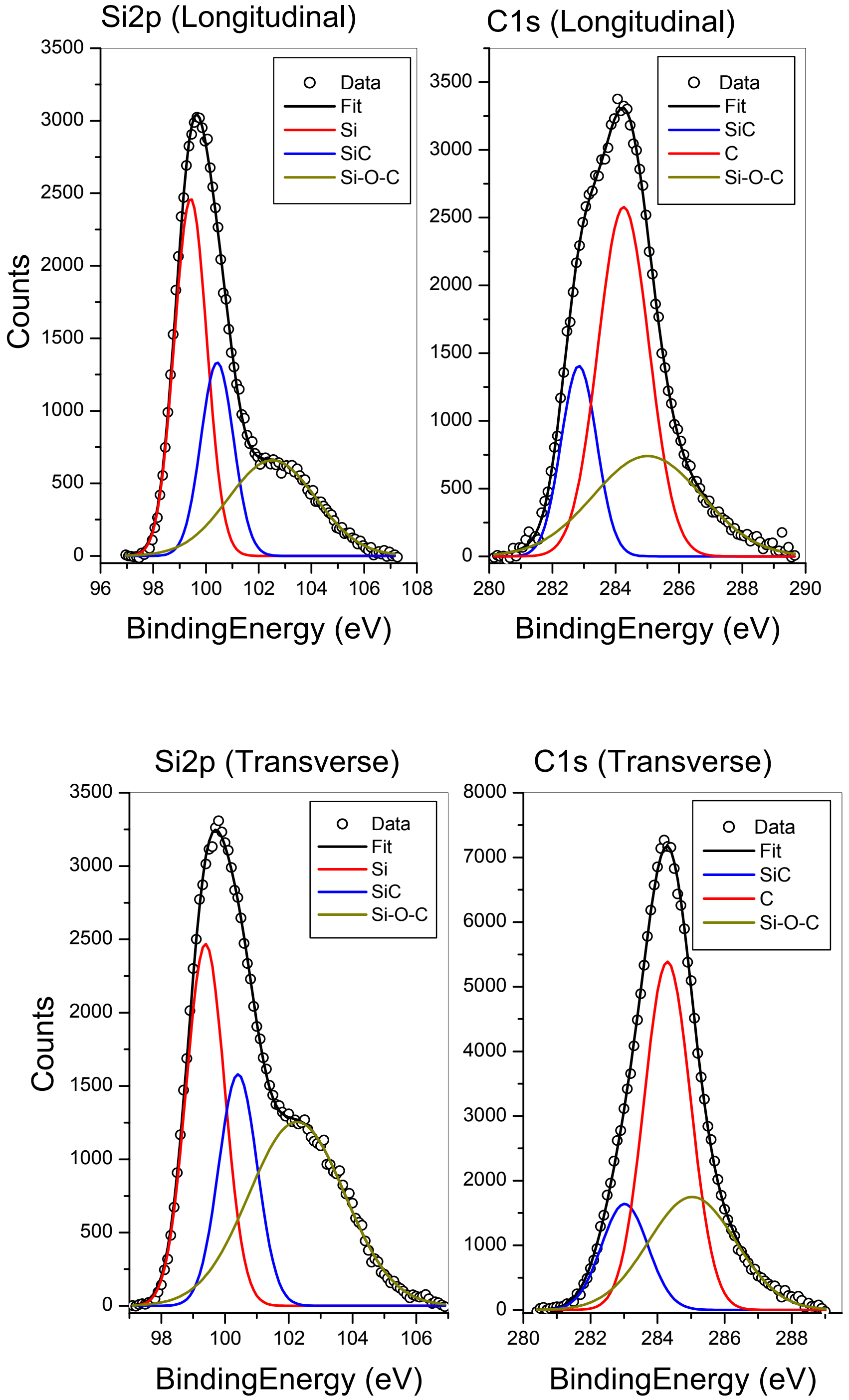

Fig.6. (color online) Si2p and C1s XPS spectra for the longitudinal and transverse directions.